\documentclass[aps,twocolumn,showpacs,prl]{revtex4-1}
\usepackage{graphicx,graphics}
\usepackage{amsmath,amssymb,amsfonts}
\usepackage{multirow}
\usepackage{hyperref}

\begin{document}
\title{Long-Range Navigation on Complex Networks using L\'evy Random Walks}
\author{A. P. Riascos}
\author{Jos\'e L. Mateos}
\affiliation{Instituto de F\'isica, Universidad Nacional Aut\'onoma de M\'exico, 
Apartado Postal 20-364, 01000 M\'exico, D.F., M\'exico}
\date{\today}

\begin{abstract} 
We introduce a strategy of navigation in undirected networks, including regular, random, and complex networks, that is inspired by L\'evy random walks, generalizing previous navigation rules. We obtained exact expressions for the stationary probability distribution, the occupation probability, the mean first passage time, and the average time to reach a node on the network. We found that the long-range navigation using the L\'evy random walk strategy, compared with the normal random walk strategy, is more efficient at reducing the time to cover the network. The dynamical effect of using the L\'evy walk strategy is to transform a large-world network into a small world. Our exact results provide a general framework that connects two important fields: L\'evy navigation strategies and dynamics on complex networks.
\end{abstract}


\maketitle
\section{I. INTRODUCTION}
Networks are ubiquitous in almost every aspect of the human endeavor, explaining the recent burst of work in this area \cite{Newman,BarabasiNature2012,Bocca,Albert}. Besides the topology of the different kinds of networks, the dynamical processes that take place on them is of utmost importance. In particular, random walks are the natural framework to study diffusion, transport, navigation, and search processes in networks, with applications in a variety of systems, such as the propagation of epidemics and traffic flow \cite{VespiBook,VespiNatPhys}, animal \cite{BES2004,Proc2006,BoyerJRSI} and human mobility \cite{Brock2006,Gonzalez2008,Boguna2009,Song2010,PRX,Simini2012}, and the dynamics on social networks \cite{Lazer2009,Moro,Latora,Aral,Holme}. 

The problem of random walks on networks has been addressed before, using a strategy of navigation that considered the motion to nearest neighbors \cite{NohRieger}.
Here we introduce a generalization of this navigation rule by considering that the transition probability is not restricted to nearest neighbors, allowing transitions that follow a power law as a function of the distance (integer number of steps) between nodes. This generalized navigation rule was inspired by the study of L\'evy flights where the random displacements $l$ obey asymptotically a power-law probability distribution of the form $P(l)\sim l^{-\alpha}$ \cite{Ralf2004}. 
\\
Random walks on networks are related to the problem of searching since one strategy of search is precisely to navigate the network, starting from a source node, using a random walk until finding a target node \cite{VespiBook}. The problem of searching and foraging has received considerable attention recently \cite{Search1,*Search2,*Search3,*Search4,ViswaBook,BeniRMP}. In particular, it has been shown that, under some general circumstances, L\'evy flights provide a better strategy to search or navigate, compared with a strategy based on Brownian motion \cite{ViswaBook,Lomholt}. For instance, in the problem of foraging by animals \cite{ViswaBook, Sims,Jager}, L\'evy strategies are rather common, as well as in the problem of human mobility and behavior \cite{Brock2006,Brown,Rhee,Chaos2011,Baron1,Baron2}. In a similar fashion, we are proposing that our L\'evy random walk navigation strategy (LRW) can be more efficient than the normal random walk strategy (NRW) to cover the network.  
\\
This generalized navigation strategy can consider some common situations encountered in real networks. For instance, in social networks one can take advantage of the full or partial knowledge of the network beyond our first acquaintances. Currently, using social network sites, one can identify the friends of your acquaintances (second-nearest neighbors) or the friends of the friends of your acquaintances (third-nearest neighbors) and so on, to search for a job, an expert opinion, etc. In this way, one can contact a second- or third-nearest neighbor of a friend directly, without the intervention of the friend. This situation corresponds to a long-range navigation on a network: a social network in this case.
\\
Thus, in this paper we are connecting two important fields: L\'evy navigation strategies and dynamics on complex networks.

We study the dynamics on an undirected network by means of a master equation that we solve exactly without any approximation. We obtain exact expressions for the stationary distribution, the random walk centrality \cite{NohRieger}, the mean first passage time (MFPT) \cite{SRedner}, and the average time to reach any node on the network. We use a formalism in terms of eigenvalues and eigenvectors of the transition probability matrix associated with the process. 

\section{II. LEVY RANDOM WALKS ON NETWORKS}

We consider an undirected connected network with $N$ nodes $i=1,...,N$, described by an adjacency matrix $\mathbf{A}$ with elements $A_{ij}=A_{ji}=1$ if there is a link between $i$, $j$ and $A_{ij}=A_{ji}=0$ otherwise. We consider the case where $A_{ii}=0$ to avoid loops on the network. The degree of the node $i$ is given by $k_i=\sum_{l=1}^N A_{il}$. Another matrix associated with the network is the distance matrix $\mathbf{D}$, with elements $d_{ij}$ that denote the integer number of steps of the shortest path connecting node $i$ to node $j$. For undirected networks $\mathbf{D}$ is a  symmetric $N\times N$ matrix. The average distance $\left\langle d\right\rangle$ scales as a power of $N$ for large-world networks, whereas $\left\langle d\right\rangle$ scales as $\log N$ for small-world networks \cite{Newman}.
\\
We start with the discrete time master equation that describes a random walker on a network \cite{Hughes}
\begin{equation}\label{master}
P_{ij}(t+1) = \sum_{m=1}^N  P_{im} (t) w_{m\rightarrow j} \ ,
\end{equation}
where $P_{ij}(t)$ is the occupation probability to find the random walker in $j$ at time $t$ starting from $i$ at $t=0$. The quantity $w_{i\rightarrow j}$ is the transition probability to move from $i$ to $j$ in the network. In the case where $i \neq j$, it is given by
\begin{equation}\label{wij}
w_{i\rightarrow j}= \frac{d_{ij}^{-\alpha}}{\sum\limits_{l \neq i} d_{il}^{-\alpha}},\\
\end{equation}
where $d_{ij}$ are the elements of the distance matrix $\mathbf{D}$ and $w_{i\rightarrow i}=0$. This transition probability represents a dynamical process where the random walker can visit not only nearest neighbors but nodes farther away in the network. However, the farther away they are, the less probable the event of hopping to that node is. The power-law decay of this probability is controlled by an exponent $\alpha$, which is a parameter in our model and varies in the interval $0\leq\alpha<\infty$. 
\\
There are two important limiting cases: In Eq. (\ref{wij}), when $\alpha \to \infty$ we obtain $w_{i\rightarrow j}=A_{ij}/k_i$, which corresponds to the normal random walk on networks, previously studied by other authors \cite{NohRieger}, describing transitions only to nearest neighbors with equal probability, that is, inversely proportional to the degree of the node. On the other hand, when $\alpha=0$, the dynamics allows the possibility of hopping with equal probability to any node on the network; in this limit, $w_{i\rightarrow j}=(1-\delta_{ij})/(N-1)$, where $\delta_{ij}$  denotes the Kronecker delta. For $0<\alpha<\infty$, the random walker could hop not only to nearest neighbors but to second-, third- and $m$-nearest neighbors with a transition probability that decays as a power law.
\\
It is worth noticing that there are navigation strategies where some links are added to a $d$-dimensional lattice with a probability proportional to $r^{-\alpha}$, where $r$ is a metric distance, generating in this way a small-world network \cite{Nature_Kleinberg,*PRL_Carmi,*PRL_Stanley,*PRL_Havlin}. However, in our case, the network is not modified, and the possibility of long-range steps is determined dynamically by $w_{i\to j}$ in Eq. (\ref{wij}).  
\\ 

Let us now solve this problem, starting with the stationary distribution. By iteration of Eq. (\ref{master}), $P_{ij}(t)$ takes the form
\begin{equation}\label{factors}
P_{ij}(t) = \sum_{j_1,\ldots,j_{t-1}} 
w_{i\rightarrow j_1} w_{j_1\rightarrow j_2} 
\cdots  
w_{j_{t-1}\rightarrow j} \, .
\end{equation}
Defining the quantity $D_i^{(\alpha)}\equiv\sum\limits_{l\neq i}d_{il}^{-\alpha}$, from Eq. (\ref{wij}), we obtain 
$w_{i\rightarrow j}=\frac{D_{j}^{(\alpha)}}{D_{i}^{(\alpha)}}w_{j\rightarrow i}$. Using this relation in Eq. (\ref{factors}), the detailed ba\-lan\-ce condition is obtained:
\begin{equation}\label{Dbalance}
   D_{i}^{(\alpha) }P_{ij}(t)=D_{j}^{(\alpha)}P_{ji}(t) \ .
\end{equation}
In order to obtain the stationary distribution, we need to take the limit of the occupation probability when the time tends to infinity. In this limiting case, the information of the initial condition is lost, and the stationary distribution gives the asymptotic probability to be in a particular node. Thus, for the stationary distribution $P_j^{\infty}=\lim_{t \to \infty}P_{ij}(t)$ \footnote{If $\lambda_N\neq -1$, $P_i^\infty=\lim_{t\to\infty}P_{ij}(t)$; if $\lambda_N= -1$ the Markovian process is cyclic and $P_i^\infty$ is interpreted as a temporal average of $P_{ij}(t)$ when $t\to\infty$.}, Eq. (\ref{Dbalance}) implies $D_{i}^{(\alpha)} P_j^{\infty}=D_{j}^{(\alpha)} P_i^{\infty}$; therefore
\begin{equation}\label{Pinf}
   P_i^{\infty}=\frac{ D_{i}^{(\alpha)}}{\sum_{l}D_{l}^{(\alpha)}} \, .
\end{equation}
Thus, we have obtained in Eq. (\ref{Pinf}) the exact expression of the stationary distribution $P_i^\infty $, which is proportional to the quantity $D_{i}^{(\alpha)}$ given by the sum of the inverse of the distances, weighted by $\alpha$, to node $i$. Equation (\ref{Pinf}) generalizes previous results, and it is valid for any undirected network. Again, we have two limiting cases: When $\alpha \to \infty$, Eq. (\ref{Pinf}) gives $P_i^\infty =\frac{k_i}{\sum_j k_j}$, which is precisely the result obtained for normal random walks \cite{VespiBook,NohRieger}; for $\alpha=0$, we obtain the expected result $P_i^\infty =\frac{1}{N}$.
\\
The quantity $D_{i}^{(\alpha)}$ can also be written as
\begin{equation}\label{Dsum}
  D_{i}^{(\alpha)} =  \sum_{n=1}^{N-1} \frac{1}{n^{\alpha}} k_i^{(n)}=k_i+ \frac{k_i^{(2)}}{2^{\alpha}}+ \frac{k_i^{(3)}}{3^{\alpha}}+\ldots   ,
\end{equation}
where $ k_i^{(n)}$ is the number of $n$-nearest neighbors of node $i$. This equation gives a more clear interpretation of the quantity $ D_{i}^{(\alpha)}$: For  node $i$, it is the sum of the first-nearest neighbors plus the second-nearest neighbors divided by $2^\alpha$ and so forth. We refer to this quantity as the \textit{long-range degree}.

We study now the random walk centrality and the MFPT. The occupation probability $P_{ij}(t)$ in Eq. (\ref{master}) can be expressed as \cite{NohRieger,Hughes}
\begin{equation}\label{EquF}
P_{ij}(t) = \delta_{t0} \delta_{ij} + \sum_{t'=0}^t   P_{jj}(t-t')  F_{ij}(t') \ ,
\end{equation}
where $F_{ij}(t)$ is the first-passage probability starting in node $i$ and finding node $j$ for the first time after $t$ steps. Using the discrete Laplace transform $\tilde{f}(s) \equiv\sum_{t=0}^\infty e^{-st} f(t)$ in Eq. (\ref{EquF}), we have  
\begin{equation}\label{LaplTransF}
\widetilde{F}_{ij} (s) = \left[\widetilde{P}_{ij}(s) - \delta_{ij}\right] / 
\widetilde{P}_{jj} (s) \ .
\end{equation}
In finite networks, the MFPT, defined as the mean number of steps taken to reach node $j$ for the first time, starting from node $i$ \cite{Hughes}, is given by $\langle T_{ij}\rangle \equiv \sum_{t=0}^{\infty} t F_{ij} (t) = -\widetilde{F}'_{ij}(0)$. Using the moments $R^{(n)}_{ij}\equiv \sum_{t=0}^{\infty} t^n ~ \{P_{ij}(t)-P_j^\infty\}$, the expansion in series of $\widetilde{P}_{ij}(s)$ is
\begin{equation}
\widetilde{P}_{ij}(s) =P_j^\infty\frac{1}{(1-e^{-s})}
+ \sum_{n=0}^\infty (-1)^n R^{(n)}_{ij} \frac{s^n}{n!} \ .
\end{equation}
Introducing this result in Eq. (\ref{LaplTransF}), we obtain for the MFPT the expression:
\begin{equation}\label{Tij}
\langle T_{ij} \rangle =\frac{1}{P_j^\infty}\left[R^{(0)}_{jj}-R^{(0)}_{ij}+\delta_{ij}\right] .
\end{equation}
In Eq. (\ref{Tij}) there are three different terms: The mean first return time $\langle T_{ii} \rangle=\frac{1}{P_i^\infty}$, the quantity  $\tau_j\equiv \frac{R_{jj}^{(0)}}{ P_j^\infty}$ which is independent of the initial node, and $\frac{R_{ij}^{(0)}}{ P_j^\infty}$. The time $\tau_i$ is interpreted as the average time needed to reach node $i$ from a randomly chosen initial node of the network. The quantity $C_i\equiv\tau_i^{-1}$ is the random walk centrality introduced in \cite{NohRieger} and gives the average speed to reach the node $i$ performing a random walk. In our case, this quantity corresponds to a generalization of the random walk centrality based on the long-range degree given in Eq. (\ref{Dsum}). 
\\
Using Eq. (\ref{Tij}), we obtain
\begin{equation}\label{Tconst}
	\langle \bar{T} \rangle \equiv\sum_{\substack{j=1\\i\neq j}}^N \langle T_{ij}\rangle P_j^\infty = \sum_{m=1}^N R_{mm}^{(0)}\, ,
\end{equation}
where the time $\langle \bar{T} \rangle$ is the average of the MFPT over the stationary distribution $P_j^\infty$ and, as shown here, is a constant independent of $i$. In the context of Markovian processes it is known as Kemeny's constant and is related to a global MFPT \cite{GMFPTKemeny,*GMFPTZhang,*GMFPTBenichou}.

In order to calculate $\tau_i$ and $\langle T_{ij} \rangle$ we need to find $P_{ij}(t)$. We start with the matrical form of the master equation $\vec{P}(t)=\vec{P}(0)\mathbf{W}^t$. The transition probability matrix $\mathbf{W}$ is a stochastic matrix with elements $W_{ij}=w_{i\to j}$, and $\vec{P}(t)$ is the probability vector at time $t$. Using Dirac's notation,
\begin{equation}\label{ProbVector}
P_{ij}(t)=\left\langle i\right|\mathbf{W}^t \left|j\right\rangle,
\end{equation}
where $\{\left|m\right\rangle \}_{m=1}^N$ represents the canonical base of $\mathbb{R}^N$. Due to the existence of the detailed balance condition, the matrix $\mathbf{W}$ can be diagonalized, and its spectrum has real eigenvalues \cite{StoMat}. We find a solution of (\ref{master}) in terms of the right eigenvectors of the stochastic matrix $\mathbf{W}$ that satisfy $\mathbf{W}\left|\phi_i\right\rangle=\lambda_i\left|\phi_i\right\rangle $ for $i=1,..,N$. The set of eigenvalues is ordered in the form $\lambda_1=1$ and $1>\lambda_2\geq..\geq\lambda_N\geq -1$. Using the right eigenvectors, we define the matrix $\mathbf{Z}$ with elements $Z_{ij}=\left\langle i|\phi_j\right\rangle$. The matrix $\mathbf{Z}$ is invertible, and a new set of vectors $\left\langle \bar{\phi}_i\right|$ is obtained by means of $Z^{-1}_{ij}=\left\langle \bar{\phi}_i |j\right\rangle $. Thus
\begin{equation}
\delta_{ij}=(\mathbf{Z}^{-1}\mathbf{Z})_{ij}=\sum_{l=1}^N \left\langle\bar{\phi}_i|l\right\rangle \left\langle l|\phi_j\right\rangle=\langle\bar{\phi}_i|\phi_j\rangle \, .
\end{equation}
Using the diagonal matrix $\Lambda \equiv \text{diag}(\lambda_1,..,\lambda_N)$, we obtain $\mathbf{W}=\mathbf{Z}\Lambda\mathbf{Z}^{-1}$. Therefore, Eq. (\ref{ProbVector}) takes the form
\begin{equation}\label{EigenPt}
   P_{ij}(t)=\left\langle i\right|\mathbf{Z}\Lambda^t\mathbf{Z}^{-1}\left|j\right\rangle=\sum_{l=1}^N\lambda_{l}^t\left\langle i|\phi_l\right\rangle \left\langle \bar{\phi}_l|j\right\rangle  \, .
\end{equation}
From Eq. (\ref{EigenPt}), $P_j^{\infty}=\left\langle j|\phi_1\right\rangle\left\langle \bar{\phi}_1|j\right\rangle$, where the result $\left\langle i|\phi_1\right\rangle=$ const was used. Using the definition of $R_{ij}^{(0)}$, we have
\begin{equation}
\label{RijSpect}
R_{ij}^{(0)}=\sum_{l=2}^N\frac{1}{1-\lambda_l}\left\langle i|\phi_l\right\rangle \left\langle\bar{\phi}_l|j\right\rangle \, .
\end{equation}
Therefore, the time $\tau_i$ is given by
\begin{equation}
\label{TauiSpect}
   \tau_i=\sum_{l=2}^N\frac{1}{1-\lambda_l}\frac{\left\langle i|\phi_l\right\rangle \left\langle\bar{\phi}_l|i\right\rangle}{ \left\langle i|\phi_1\right\rangle\left\langle \bar{\phi}_1|i\right\rangle}\, ,
\end{equation}
and for $i \neq j$ in Eq. (\ref{Tij}), the MFPT $\left\langle T_{ij}\right\rangle$ is
\begin{equation}\label{TijSpect}
\left\langle T_{ij}\right\rangle
=\sum_{l=2}^N\frac{1}{1-\lambda_l}\frac{\left\langle j|\phi_l\right\rangle \left\langle\bar{\phi}_l|j\right\rangle-\left\langle i|\phi_l\right\rangle \left\langle\bar{\phi}_l|j\right\rangle}{\left\langle j|\phi_1\right\rangle\left\langle \bar{\phi}_1|j\right\rangle}\, .
\end{equation}
Finally, using (\ref{Tconst}) and (\ref{RijSpect}), we obtain
\begin{equation}\label{TconstSpect}
\langle \bar{T} \rangle=\sum_{m=1}^N\sum_{l=2}^N \frac{1}{1-\lambda_l} \left\langle\bar{\phi}_l|m\right\rangle \left\langle m|\phi_l\right\rangle =\sum_{l=2}^N \frac{1}{1-\lambda_l}\, .
\end{equation}
Therefore, we have obtained exact expressions for the occupation probability $P_{ij}(t)$, the stationary distribution $P^\infty_i$, the time $\tau_i$, the MFPT $\left\langle T_{ij}\right\rangle$, and the time $\langle \bar{T} \rangle$ in terms of the spectrum and the left and right eigenvectors of $\mathbf{W}$. Notice that the time $\langle \bar{T} \rangle$ is a constant that can be calculated using only the spectrum of eigenvalues of $\mathbf{W}$.
\begin{figure}[!t]
\begin{center}
\includegraphics[width=0.46\textwidth]{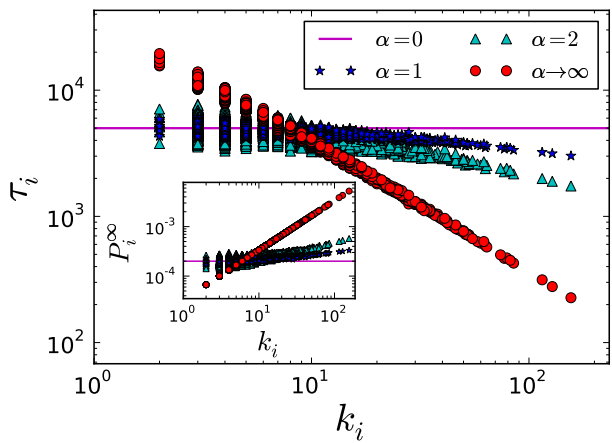}
\end{center}
\vspace{-6mm}
\caption{\label{fig:Fig1} (Color online) $\tau_i$ vs $k_i$ for a scale-free network with $N=5000$ nodes, using three values of the exponent $\alpha$. The inset depicts $P_i^{\infty}$ vs $k_i$, and the solid line indicates the limiting case $\alpha=0$.}
\end{figure} 

In order to analyze the navigation of a L\'evy random walker on the network, we introduced another global time that is simply the average of the quantity $\tau_i$ over all the nodes on the network, defined as $\tau\equiv\frac{1}{N}\sum_{m=1}^N\tau_m$. This global time $\tau$ gives the average number of steps needed to reach any node on the network independently of the initial condition. In the particular, yet important, case of regular networks (such as lattices and complete graphs) where the stationary probability is a constant given by $P_i^\infty=1/N$, we obtain that both global times $\langle \bar{T} \rangle$ and $\tau$ are the same, $\tau=\langle\bar{T}\rangle$. For example, when $\alpha=0$, the matrix $\mathbf{W}$ has eigenvalues $\lambda_1=1$ and $\lambda_j=-\frac{1}{N-1}$ for $j=2,3,\ldots ,N$, and Eq. (\ref{TconstSpect}) gives $\tau=\frac{(N-1)^2}{N}$. 

\begin{figure}[!b]
\begin{center}
\includegraphics[width=0.47\textwidth]{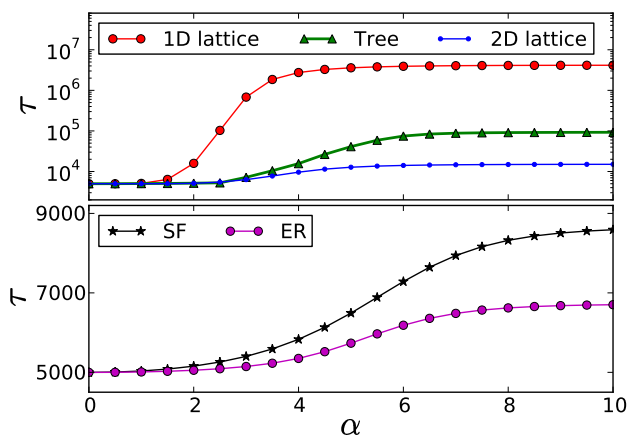}
\end{center}
\vspace{-6mm}
\caption{\label{fig:Fig2} (Color online) The time $\tau$ vs $\alpha$ for different networks with $N=5000$. (top) The results for a 1D lattice, a 2D square lattice ($50\times100$), and a random tree (network without loops); we used periodic boundary conditions in both lattices. (bottom) The results for a SF network and an ER network at the percolation threshold.}
\end{figure}

\section{III. ANALYSIS OF THE RESULTS}

In what follows we use the exact results obtained by our matrix formalism, given in Eqs. (\ref{EigenPt})-(\ref{TconstSpect}), to calculate the corresponding quantities. It is important to stress that the results in Figs. \ref{fig:Fig1} and \ref{fig:Fig2} are not numerical but exact calculations using the eigenvalues and eigenvectors of the matrix $\mathbf{W}$.
In Fig. \ref{fig:Fig1} we show the quantity $\tau_i$, which represents the average time needed to reach node $i$ from any node in the network. We use in Fig. \ref{fig:Fig1} a scale-free network of the Barab\'asi-Albert (BA) type, in which each node has a degree that follows asymptotically a power-law distribution $p(k) \sim k^{-\beta}$ \cite{NetBA}. We show three cases with different values for the exponent $\alpha$ in Eq. (\ref{wij}), that correspond to three different navigation rules. The cases $\alpha = 1, 2$ correspond to a LRW, and the limiting case $\alpha \to \infty$ corresponds to a NRW \cite{NohRieger}. In the inset we show the stationary distribution $P_i^\infty$ for different values of degree $k_i$. From Fig. \ref{fig:Fig1} we can conclude that (1) according to the inset, using LRW, it is more probable to reach the nodes with a small degree (which are the majority of nodes on the network), compared with NRW, and (2) for nodes with a degree lower than a critical degree, which is the vast majority, LRW can diminish the time to cover most of the network, compared with NRW. On the other hand, the time to reach those nodes with a high degree is lower using NRW than LRW since LRW covers more uniformly the entire network, despite the degree of the nodes. Therefore, if the purpose is to cover the vast majority of the network, it is better to use a LRW strategy, but if the goal is to reach the few highly connected nodes, it is better to use a NRW.  

In Fig. \ref{fig:Fig2}, we show $\tau$ vs $\alpha$ for five different kinds of networks using our exact results. In regular networks [one-dimensional (1D) and two-dimensional (2D) lattices] $\tau$ is calculated using Eq. (\ref{TconstSpect}), and for the other three networks $\tau$ is obtained by averaging the quantities given by (\ref{TauiSpect}). The top panel of Fig. \ref{fig:Fig2} shows that for large-world networks, such as lattices and trees (networks without loops \cite{Newman}), the LRW strategy navigates the network more efficiently. That is, for smaller values of $\alpha$ the average number of steps tends to the value $(N-1)^2/N$, whereas for larger values of $\alpha$ (corresponding to NRW) the number of steps can be one or two orders of magnitude larger. In the bottom panel we notice that even for small-world networks, such as scale-free (SF) and Erd\H{o}s-R\'enyi (ER) \cite{VespiBook} networks, the number of steps is larger for NRW than for LRW. Thus, the LRW strategy reduces the global time $\tau$ compared with the NRW strategy, transforming dynamically a large-world network into a small world.

It is worth noticing that the walker can in principle visit any node of the network in a single step, but with a decreasing probability depending on the exponent $\alpha$. The dynamics involve small, intermediate and large steps, drawn from a probability distribution, given by Eq. (\ref{wij}), that decays as a power law. Thus, the dynamics is more complex than in a complete graph, where every node is connected to every other node; however, as discussed before, in the limiting case when $\alpha$ tends to zero, the transition probability tends to a constant value given by $w_{ij} = (1-\delta_{ij})/(N-1)$, where $N$ is the number of nodes on the network. In this limiting case, we arrive at a situation similar to a complete graph where each node on the network is connected to every other node. We show in Fig. \ref{fig:Fig2}, that the case $\alpha = 0$ is optimum in the sense that the average global time is minimum; however, even for values of $\alpha \approx 1$, the difference between the LRW strategy and the dynamics for $\alpha = 0$ is negligible. This implies that the overall effect of navigating using L\'evy random walks can be as effective as navigating on a complete graph.
\begin{figure}[!t]
\begin{center}
\includegraphics*[width=0.47\textwidth]{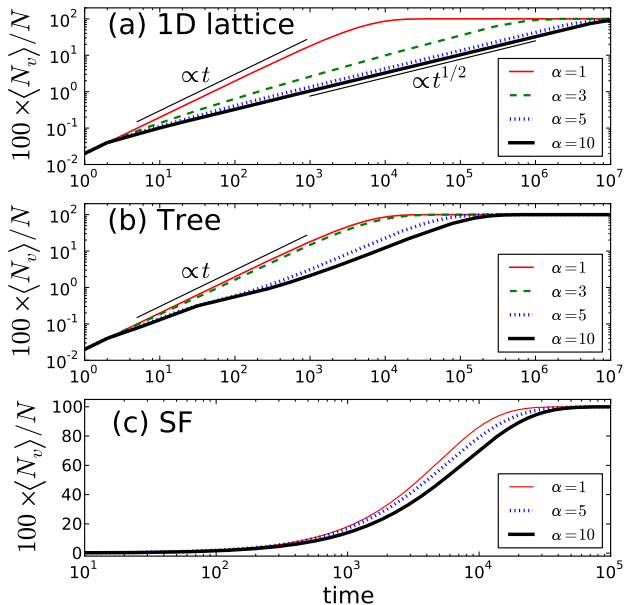}
\end{center}
\vspace{-6mm}
\caption{\label{fig:Fig3} (Color online) Monte Carlo simulation of the number of distinct visited sites $N_v$ vs time in a network with $N=5000$. (a) A 1D lattice, (b) a random tree, and (c) a scale-free network. Each curve is obtained from the average of $1000$ realizations of the random walker.}
\end{figure}

These results are confirmed in Fig. \ref{fig:Fig3} where we show now Monte Carlo simulations for the percentage of different visited sites $N_v$ in the network, as a function of time, for different values of the exponent $\alpha$. Small values ($\alpha = 1,3$) correspond to LRW and large values ($\alpha = 10$) to NRW. In Fig. \ref{fig:Fig3}(a), we depict the results for a 1D lattice, showing that the number of visited nodes for NRW grows diffusively, whereas for LRW grows ballistically. In Fig. \ref{fig:Fig3}(b), we show the results for trees and we see that once again the  LRW explores the network faster than the NRW. Finally, in Fig. \ref{fig:Fig3}(c) we show that even for scale-free networks there is some advantage in exploring the network using the LRW strategy than the NRW strategy. Even though the difference is smaller, we notice that we can cover the network faster using LRW than using NRW. 

\section{CONCLUSIONS}

In summary, we have introduced a new strategy of navigation in general undirected networks, including complex networks, inspired by L\'evy random walks, that generalized previous navigation rules. We obtained exact expressions, using a matrix formalism, for the stationary probability distribution, the occupation probability distribution, the mean first passage time and the average time to reach a node in any undirected finite network. We found that using the L\'evy random walk navigation strategy we cover more efficiently the network in comparison with the normal random walk strategy. For small-world networks we obtained that the average time to reach any node on the network is smaller than the time required using the normal random walk strategy. For large-world networks, this difference can be one or two orders of magnitude. Additionally, we found that for large-world networks, like trees or lattices, the L\'evy navigation strategy can induce dynamically the small-world effect, thus transforming a large-world network into a small world. Finally, our exact results provide a general formalism that connects two important fields: L\'evy navigation strategies and dynamics on complex networks. 

\begin{acknowledgments}
J.L.M. thanks M. Shlesinger and A.-L. Barab\'asi for useful comments. A.P.R. acknowledges support from CONACYT M\'exico.
\end{acknowledgments}
\providecommand{\noopsort}[1]{}\providecommand{\singleletter}[1]{#1}%
\end{document}